\documentclass[twocolumn,showpacs,amsmath,amssymb]{revtex4}

\usepackage{graphicx} 
\usepackage{dcolumn}  
\usepackage{bm}       
\usepackage{verbatim}       

\begin{document}

\title{Kinetic Pathways of the DNA Melting Transition}

\author{A. Santos$^1$ and W. Klein$^2$}

\affiliation{$^1$Physics Department, Gustavus Adolphus College, St. Peter, MN 56082}
\affiliation{$^2$Department of Physics, Boston University, Boston, MA 02215}

\date{\today}

\begin{abstract} 
We investigate kinetic pathways of the DNA melting transition using variable-range versions of the  Poland-Scheraga (PS) and  Peyrard-Dauxois-Bishop (PDB) models of DNA.  
In the PS model, we construct a $\phi^4$-field theory to calculate the critical droplet profile, the initial growth modes, and the exponent $\gamma$ characterizing the divergence of the susceptibility near the spinodal. 
In the PDB model, we use a mean field analysis to calculate $\gamma$.  
We compare these theoretical results with Monte Carlo and Brownian dynamic simulations on the PS and PDB models, respectively.
We find that by increasing the range of interaction, the system can be brought close to a pseudospinodal, and that in this region the nucleating droplet is diffuse in  contrast to the compact droplets predicted by classical nucleation theory.
\end{abstract}


\maketitle
\section{Introduction}
The DNA melting transition is an interesting theoretical problem because a quantitative understanding of its mechanism may provide insight into how biological enzymes physically interact with DNA~\cite{Wartell}.   
While much theoretical work has examined the role of large nonlinear excitations as a precursor to melting ~\cite{Englander,Dauxois, Dauxois2}, surprisingly little focus has been paid to kinetics of the transition itself. 
In this article, we use classical and spinodal nucleation theory in conjuncton with simulations to study the kinetics of melting in both short- and long-range versions of the Peyrard-Dauxois-Bishop (PDB)~\cite{Dauxois,Dauxois2} and Poland-Scheraga (PS)~\cite{Wartell,Poland} models of DNA.

Nucleation plays an important role in many systems undergoing a phase transition~\cite{Zettlemoyer}.  
In homogeneous systems, a droplet forms from a spontaneous fluctuation and grows into the stable phase.  
Before nucleation, the system is trapped in a metastable well in a free energy landscape.  
The system samples the phase space of this metastable well until a fluctuation drives the system to the top of a free energy barrier that separates the stable and metastable wells.  
At this point, the droplet is referred to as critical and the system is equally likely to nucleate or return to the metastable well.  
DNA is believed to undergo such a process because its sharp melting curve indicates a first-order phase transition~\cite{Wartell} and because hysteresis in the melting curve suggests the existence metastable states~\cite{Hoff,Michel,Yabuki}. 
In addition, possible nucleation bubbles have been observed via electron microscopy~\cite{Pavlov}. 

There is experimental evidence for the presence of long-range (LR) interactions in DNA~\cite{Wartell,Wartell2,Wartell3,Burd,Gotoh,Werntges, Coman}.  
Telestability experiments on block copolymers show cooperativity effects over at least 10-15 base pairs (bp)~\cite{Wartell2} while differential melting curves with multistep behavior show cooperatively melted regions of 100-350 bps~\cite{Wartell}.  
In low salt concentration, cooperatively melted regions can be as a large as a few thousand bps~\cite{Gotoh}.
Both experimental and  molecular dynamics studies suggest that bases beyond nearest-neighbor can effect the  enthalpic change that arises from the opening of a given bp ~\cite{Coman, Varnai}.   
In addition, it is known that nearest-neighbor (NN) PS models underestimate the probability of single bp opening~\cite{Blossey}.  
Only recently have researchers begun examining how LR interactions affect the dynamics of DNA models \cite{Rapti}.  
To consider the effects of the interaction range on the kinetic pathway, we simulate the Peyrard-Dauxois-Bishop (PDB)~\cite{Dauxois, Dauxois2} and PS~\cite{Poland,Blossey} models with both NN and LR interactions.

It is well known that LR systems undergoing a phase transition can be quenched into metastable states near a pseudospinodal \cite{Laradji, Herrmann, Klein}.  
In the mean field (MF) limit $R\rightarrow\infty$, this pseudospinodal becomes a well-defined spinodal~\cite{Klein}, which is the limit of metastability.  
The barrier to nucleation for a $d$-dimensional system scales with the interaction range $R$ as $R^d\Delta h^{3/2-d/4}$, where $\Delta h$ is the distance away from the spinodal~\cite{Klein}.
As such, the nucleation rate is much smaller for LR systems, and, practically, one must get very close to the spinodal to observe nucleation. 
Nucleation near a spinodal or pseudospinodal is similar to classical nucleation near the coexistence curve in that droplets become critical by reaching the top of a free energy barrier and then either grow 
or decay with equal probability.  
Unlike classical nucleation, which is initiated by compact droplets that resemble the stable phase~\cite{Langer}, spinodal nucleation is characterized by the formation of diffuse fractal-like droplets 
whose amplitude differs little from the metastable background~\cite{Unger, Unger2, Gould}.  
In addition, the growth modes of classical droplets lie on the droplet surface, while spinodal droplets grow from their center.  
For these reasons, the inclusion or exclusion of LR interactions in DNA models makes a substantial difference in regard to the character of nucleation that can be observed.

In addition to changing the qualitative shape of the critical droplet, moving the system toward a pseudospinodal also causes the isothermal susceptibility $\chi$ to diverge.  
This effect has been observed in supercooled water~\cite{Speedy}, and its measurement may be a useful method for determining whether or not nucleation in real DNA exhibits spinodal effects.
While there is inherent fuzziness in the definition of the pseudospinodal in real systems~\cite{Gulbahce2}, practically one can determine both its location and the exponent $\gamma$ characterizing the divergence if the metastable lifetime is longer than the measurement time of $\chi$.
If a spinodal exists for DNA, it should  in priniciple be possible to measure both the spinodal temperature $T_s$ and $\gamma$. 

Finally, it is important to note that biological DNA is an intrinsically heterogeneous system because of its pseudo-random sequence of bases.
In this work, we concentrate on homogeneous nucleation, which is purely initiated by spontaneous fluctuations, rather than in heterogeneous nucleation, where nucleation is aided by the presence of boundaries, defects, or other impurities~\cite{Fletcher}.
However, since heterogeneous sequences of bases are clearly important biologically, we simulate random base pair sequences to determine what effects the inhomogeneities have on the kinetic pathways of the transition.

This paper is organized as follows.  Sec.~II provides a detailed description of the modified PDB and PS models.  In Sec.~III, we postulate a phenomenological free energy for the PS model and use 
this to calculate the shape of the droplet profile, initial growth modes, and the susceptibility exponent.  In Sec.~IV, we provide a MF calculation of the susceptibility
exponent in the PDB model.  In Sec.~V, we describe the results of our simulations of long and short range PDB and PS models including evidence for nucleation, the shape of droplet profiles and growth 
modes, and the calculation of the spinodal exponent.  We summarize and discuss the practical implications our results in Sec.~VI.

\section{PDB and PS Models}

We modified both the PDB and PS models to include long range interactions.  
In the modified PDB model, the Hamiltonian is given by 
\begin{equation}
H_{PDB} =  \sum_{i=1}^N \frac{m\dot{y}_i^2}{2}+V(\{y_i\}), 
\end{equation}
where the state of each bp is specified by its separation $y_i$ and its time derivative $\dot{y}_i$.  The first term is the kinetic energy of bps with combined mass $m=300$ amu and the second term is a potential given by 
\begin{equation}
\label{vpb}
    \begin{split}
V(\{y_i\}) & =D_i(e^{-ay_i}-1)^2+ \sum_{j=i-R}^{i-1}W(y_i,y_j) 
    \end{split}
\end{equation}

The first term in Eq.~\ref{vpb} is an on-site Morse potential describing the net attraction between strands due to a combination of hydrogen bonding, solvent interactions, and the repulsion of negatively charged phosphate groups~\cite{Dauxois, Dauxois2, Morse}. 
For the homogeneous case the dissociation energy $D_i$ is treated as a constant $D=0.04$ eV while for the heterogeneoues case it is given by $D_{AT}=0.032$ eV and $D_{GC}=0.048$ eV for A-T and C-G bps, respectively.
The inverse well width is given by $a=4.45$ \AA$^{-1}$.  

There is ample experimental evidence that suggest LR interactions play a significant role in the melting of real DNA~\cite{Wartell,Wartell2,Wartell3,Burd,Gotoh,Werntges, Coman}.
While the origin and precise nature of these interactions has yet to be characterized, we may still glean some information about the qualitative effects  LR interactions have on the models.
In the PDB model, the interaction term $W(y_n,y_m)$  provides an anharmonic potential between bases and may be written
\begin{equation}
W(y_n,y_m)=\frac{1}{2}K\big(1+\rho e^{-\alpha (y_n+y_m)} \big)(y_n-y_m)^2,
\end{equation}
where $K=0.06$ eV/\AA, $\alpha=0.35$ \AA$^{-1}$, and $\rho=0.5$.  
In the original PDB model, this term represented the stacking interaction, which is purely NN such that $R=1$~\cite{Dauxois,Dauxois2}.
Here, we interpret this interaction term broadly as an effective potential originating from several effects including stacking, backbone flexibility, hydrophilic/hydrophobic interactions, and any LR interactions.
Under this interpretation, we allow this interaction to extend to a range $R\geq 1$ and examine the effect LR interactions have on the kinetic pathways of the transition.  
This inclusion of LR interactions is similar to that used by Rapti in the Peyrard-Bishop model \cite{Rapti}.

In the modified PS model, statistical weights are given to bound and unbound segments. 
A bound segment is energetically favored because of hydrogen bonding and stacking interactions.  
Unbound segments are entropically favored because single stranded loops have a much shorter persistence length allowing them to sample a larger configuration of phase space.  
Our LR PS model can be described by 
the Hamiltonian
\begin{equation}
\label{psham}
    \begin{split}
   	H_{\rm PS}  &= -E_{0,i}\sum_{i=1}^N\Big(\frac{1-\sigma_i}{2} \Big) \\
        &-\frac{K_0}{R}\sum_{i=1}^N \Big(\frac{1-\sigma_i}{2} \Big)\sum_{j=i-R}^{i-1} \Big(\frac{1-\sigma_j}{2} \Big)\\
	&-T\sum_{\mbox{\scriptsize{loops}}}\ln \Big(\Omega \frac{s^l}{l^c} \Big),
    \end{split}
\end{equation}
where $\sigma_i=-1$ and $\sigma_i=+1$ represent bound and open bps, respectively.  
Here, $E_{0,i}$ represents the binding energy which is assumed to be the same for all bps in the homogeneous case.
In the second term, the parameter $Ko$ represents the interaction between adjacent base pairs.
This term has been added so that the effects of LR interactions in the model can be studied. 
In order to have roughly consistent parameters between the two models, we have set $E_{0,i}=D_i$ and $K_O=K(1+\rho)$.

The final term in Eq.~\ref{psham} represents an effective potential due to entropic effects caused by to loops of size $l$.  
Here, we have chosen $s=74.4$ to give a biologically relevant melting temperature $T=350$ K.  
The exponent $c=2.15$ is chosen consistent with simulation results on self-avoiding random walk loops ~\cite{Kafri, Blossey}.  
Choosing $c>2$ ensures that the melting transition will be first order.
The cooperativity $\Omega=0.3$ is chosen larger than values published elsewhere so that small loops may be observed ~\cite{Blossey}.

It might be argued that we have not motivated our choice for the functional form of the LR interactions. 
We have chosen this form two reasons. 
First, the functional forms introduced in the original models have several desirable features that we would like to keep in the LR models. 
For example, the nonlinear stacking interaction in the PDB model produces a sharper transition than a simple harmonic interaction by making separated strands less rigid than bonded strands. 
Second, there is theoretical evidence that the detailed form of the LR interaction does not play an important role in determining the quality of the physics. 
For example, simple experiments predict that the interaction matrix in the Rundle-Jackson-Brown earthquake model should decay as $1/r^3$, but it has been shown that the essential long wavelength physics is captured the model by a much simpler mean-field formulation \cite{Martins}. 
As we show below, the PDB and PS models, which have vastly different functional forms, produce the same qualitative characteristics when the interactions are extended to LR. 
For these reasons, we do not believe the exact functional form of the LR interaction will significantly impact the qualitative features we observe in nucleation. 
Since our goal is purely to describe the qualitative differences between DNA melting with and without LR interactions, the functional form we have chosen is perfectly suitable. 
Moreover, since the origin of LR interactions is still unknown, an effective potential provides the most accurate description that can be obtained at present.

\section{$\phi^4$-Field Theory in the PS Model}

Without the entropic term describing the degeneracy of loop configurations, the PS Hamiltonian of Eq.~\ref{psham} is isomorphic to the Ising model in the lattice gas representation.
This suggests it may be possible to convert the PS model into a $\phi^4$-field theory in the same way as the Ising model.  
For this reason, we postulate a Landau-Ginzburg-Wilson (LGW) free energy functional of the form
\begin{equation}
\label{LGW}
F(\phi)=\int dr[R^2(\nabla \phi)^2/2+\kappa \phi^2+h \phi -TS(\phi)].
\end{equation}
Here, $\phi(r)$ is the coarse-grained magnetization 
\begin{equation}
\phi(r)=\frac{1}{L_{CG}}\sum_{i\in L_{CG}}\sigma_i,
\end{equation}
and $L_{CG}$ is the size of the coarse-grained region.  
From this definition, it is clear that $\phi(r)\in[-1,1]$.
Here, the parameter $h=L_{CG}(E_0+K_0)/2$  is the energy associated with flipping a coarse-grained region against the direction of an Ising-like magnetic field.
The parameter $\kappa$ sets the critical temperature of the field theory. 
For convenience, we chose $\kappa=K_0$ to match the microscopic model.
The exact numerical choice of parameters is not expected to change the qualitative results.
The first three terms in the integrand of Eq.~\ref{LGW} describe the energetic contributions from the PS model while the last term describes the entropic contributions arising both from the 
coarse-graining and from the degeneracy associated with unbound loops. 

In principle, the entropy $S(\phi)$ in Eq.~\ref{LGW} can be calculated in a similar way as the entropy that arises when coarse-graining an Ising model to obtain a field theory.
The difficulty of calculating this term {\it a priori} stems from the entropy's dependence on the distribution of loop sizes.
To circumvent this problem, we assume the entropy $S(\phi)$ of a given coarse-grained region is equal to the average entropy of all loop distributions consistent with the magnetization $\phi$ of the region.
This approximation is expected to be reasonable when the loop sizes are smaller than $L_{CG}$.  
For simplicity, we choose a particular coarse-graining size $L_{CG}=16$ and enumerate the $2^{16}$ possible combinations of spins.  
For each combination, we calculate $\phi$ and the loop entropy $\sum_{loops} \log(\Omega s^l/l^c)$ of the configuration.  
We then calculate the average entropy for a given $\phi$,
\begin{equation}
S(\phi)=\left< \sum_{loops} \log(\Omega s^l/l^c) \right>.
\end{equation}
This numerical determination of $S(\phi)$ can be fit to a fourth order polynomial
\begin{equation}
\label{entrop}
S(\phi)=b_1\phi+b_2\phi^2+b_3\phi^3+b_4\phi^4,
\end{equation}
where the coefficients are given by $b_1=27.5$, $b_2=7.06$, $b_3=3.83$, and $b_4=-0.701$.

Combining Eq.~\ref{LGW} and Eq.~\ref{entrop}, we now write our LGW Hamiltonian as
\begin{equation}
F(\phi)=   \int dr\left[\frac{R^2(\nabla \phi)^2}{2} +f(\phi)\right],
\end{equation} 
where 
\begin{equation}
f(\phi) =-T b_4\phi^4    -Tb_3\phi^3     -(T b_2-\kappa) \phi^2      +(h-T b_1)\phi  .
\end{equation}

\begin{figure}[t]
\begin{center}
\includegraphics[width=9.cm,height=6.0cm]{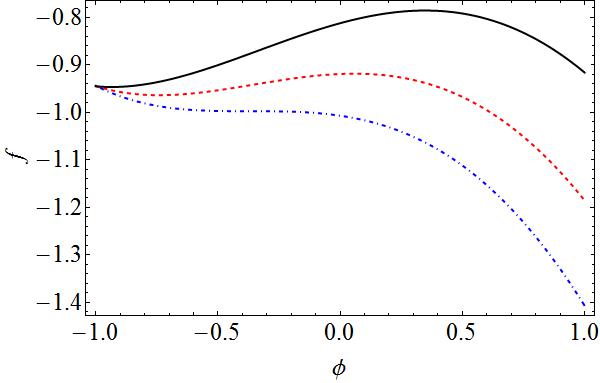}
\caption{\label{fig:freee} (color online).  Plots of the free energy density $f$ vs.~$\phi$ for various $T$.  Plots are shown for $T=350$~K (black, solid), $T=410$~K (red, dashed), and $T=471.1$~K (blue, dash-dotted).  }
\end{center}
\end{figure}
In Fig.~\ref{fig:freee}, we plot the free energy density $f(\phi)$ for $T=350$~K, $T=410$~K, and $T=471.1$~K corresponding roughly to the melting temperature $T_m$, an intermediate temperature, 
and the spinodal temperature $T_s$.
In this plot, $\phi$ is restricted to lie within a range $-1\le\phi\le 1$, consistent with its definition.
At $T$=350~K, we see that $f(\phi)$ has two wells at $\phi\approx 1$ and $\phi\approx -1$ corresponding to the open and bound states, respectively.
Since the melting temperature should lie on the coexistence curve, one would expect these wells to  be  the same depth, rather than having a deeper bound state well as show in the figure.
This discrepency arises because of the assumption of small loops and will be considered shortly.
Near the spinodal at $T=471.1$ K, the metastable well has almost vanished.
As suggested in the plot of $f(\phi)$ for $T=410$~K, $f(\phi)$ evolves continuously with a gradual disappearance of the metastable well as the spinodal temperature is approached.

From the LGW free energy, we calculate the shape of the critical droplet.  
The critical droplet is a saddle point in the free energy landscape satisfying the Euler-Lagrange equation:
\begin{equation}
\label{el}
\left.\frac{\delta F}{\delta \phi}\right|_{\bar{\phi}}=     -R^2\frac{d^2\bar{\phi}}{dr^2}      +\left.\frac{\delta f}{\delta\phi}\right|_{\bar{\phi}}=0,
\end{equation}
There are two solutions to Eq.~\ref{el} that are independent of $r$.  
One of these represents the metastable bound state $\phi=\phi_{MS}$ and the other, which would correspond to the stable unbound state, appears at $\phi>1$ and is unphysical.  
The actual stable unbound state occurs at $\phi=1$.
A spatially nonconstant solution $\bar{\phi}(r)$ with the boundary condition that $\bar{\phi}(\infty)=\phi_{MS}$ represents a fluctuation away from the metastable well.  
This fluctutation is the critical droplet.

\begin{figure}[t]
\begin{center}
\includegraphics[width=8.cm,height=9.5 cm]{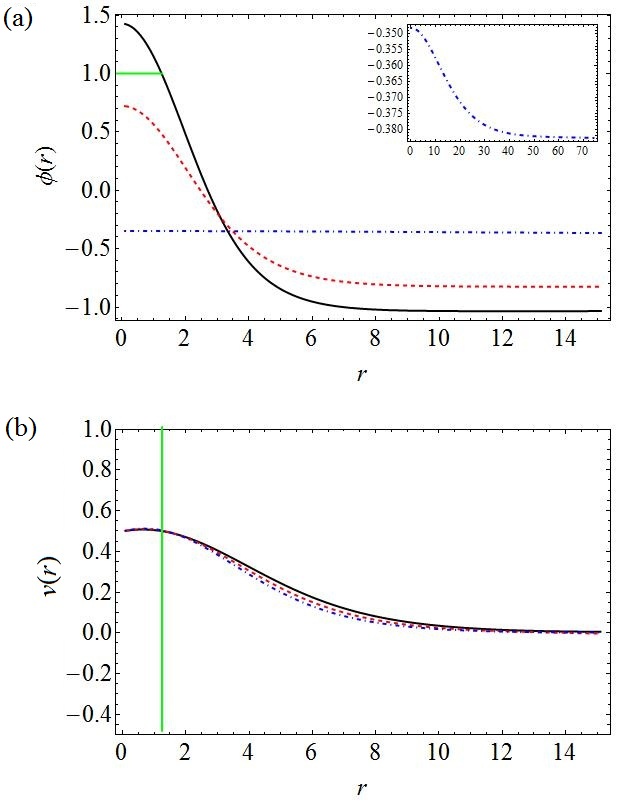}
\caption{\label{fig:PHI4DROPLETS}   (color online).  (a) Critical droplet profiles are shown for $T=350$ K (black, solid), $T=410$ K (red, dashed), and $T=471.1$ K (blue, dash-dotted).  A green line has been added to show the unphysical region where $\phi>1$.
(b) The initial growth mode $v(r)$ for each temperature of part (a).  A vertical green line has been added to denote the region where the assumption of small droplets clearly fails when $T$ is small.}
\end{center}
\end{figure}

Eq.~\ref{el} is analogous to the equation of motion for a particle in a potential $V(\phi)=-f(\phi)$ with $r$ representing time~\cite{Langer}.  
In this analog, the boundary condition $\frac{d\bar{\phi}}{dr}=0$ at $r=0$ corresponds to the particle having no initial velocity.  
The particle starts up the side of the larger hill (stable minimum) and rolls off until finally coming to rest on top of the smaller hill (metastable minimum.)  
This equation of motion was solved numerically using a fourth-order Runge-Kutta method for $T=350$~K, 410 K, and  471.1 K.
The resulting nucleating droplet profiles are shown in Fig.~\ref{fig:PHI4DROPLETS}a.  

A green line has been added to the figure to emphasize where $\phi>1$.
 At $T=350$~K, the peak of the predicted critical droplet lies above this line. 
This unphysical result is due to the assumption that loops are much smaller than the coarse-graining size $L_{CG}=16$.
Our method of calculating $S(\phi)$ undercounts the entropic contribution of loops with size $l>16$, because each coarse-grained region's entropy is calculated individually without regard to loops that may extend into the next coarse-grained cell.
This discrepency  is significant near the coexistence curve where large compact droplets are expected to nucleate the system.
Undercounting large loops increases the predicted value of the free energy for the unbound state, which can be obeserved in the plot of $f(\phi)$ in Fig. ~\ref{fig:freee}.
At the melting temperature, the free energy of unbound state should be identical to that of the bound state, but the theory predicts a significantly higher free energy for the dissociated state.
Since the magnetization is greater than one for a finite region around the center of the droplet, the actual critical droplet should be compact in this region, and the free energy of this actual compact droplet will be lower than the free energy predicted by the field theory which undercounts the entropic contribution.

Away from the coexistence curve, the field theory is a more accurate description because nucleating droplets are no longer large and compact.
From the shape of the droplet near at $T$=471.1 K  (Fig.~\ref{fig:PHI4DROPLETS} inset), this is appears to be the case.
If one notes the change in scale for the droplet near the spinodal, it is clear that the droplet must be diffuse since its amplitude differs little from the metastable background. 
At $T=410$ K, (Fig.~\ref{fig:PHI4DROPLETS}b), the droplet  is intermediate between spinodal and classical, consistent with previous results~\cite{Klein,Unger2}.

The critical droplet is perched atop a saddle point in the free energy landscape.
During the initial growth away from this point, the system rolls off the saddle along the path of steepest descent.  
If we write fluctuations to the critical droplet as $\phi(r)=\bar{\phi}(r)+v(r)$, then in the neighborhood of the saddle, the LGW free energy can 
written as $F(\phi)=F(\bar{\phi})+F''(v)$, where
\begin{equation}
F''(v)=   \int dr\left[\frac{R^2(\nabla v)^2}{2} +\left.\frac{\delta^2 f}{\delta\phi^2}\right|_{\bar{\phi}} v^2         \right].
\end{equation} 
The normal modes of the system near the critical droplet configuration are solutions to the Schr\"{o}dinger equation
\begin{equation}
\left[-R^2/2\frac{d^2}{dr^2} +\left.\frac{\delta^2 f}{\delta\phi^2}\right|_{\bar{\phi}}          \right]v_n(r)=w_n v_n(r).
\end{equation} 
There is one negative eigenvalue for this equation corresponding to an instability. 
This initial growth mode increases exponentially with time.  

For each of the temperatures listed above, the growth modes were calculated numerically using the shooting method and are depicted in Fig.~\ref{fig:PHI4DROPLETS}b.  
As before, the field theory works poorly near the coexistence curve where droplets are large and compact. 
A green line has been added to denote where the $T=350$ K droplet crosses into the unphysical region in which $\phi>1$.
In this region, the $T$=350 K droplet is expected to be compact.
From Fig.~\ref{fig:PHI4DROPLETS}b, one can  see that the theory predicts maximum growth for a $T$=350 K droplet occurs at the center of the droplet. 
A compact droplet cannot grow from its center because this region is already in the stable phase, hence $v(r)$ must be zero for small $r$.
Since $v(r)=0$ for small $r$, the maximum of the growth mode must lie on the surface of the droplet  when the system is near the coexistence curve.
Away from the coexistence curve, the theoretical results are more accurate since the droplets are diffuse.
From the figure, one can see the predicted position of maximum growth remains at the center of the droplet as the spinodal is approached. 
Unlike droplets that form near the coexistence curve, droplets that form near the spinodal are diffuse and can grow from their center. 
This agrees with previous results on nucleation in Ising models~\cite{Unger}.

\begin{figure}[t]
\begin{center}
\includegraphics[width=8.cm,height=9.5cm]{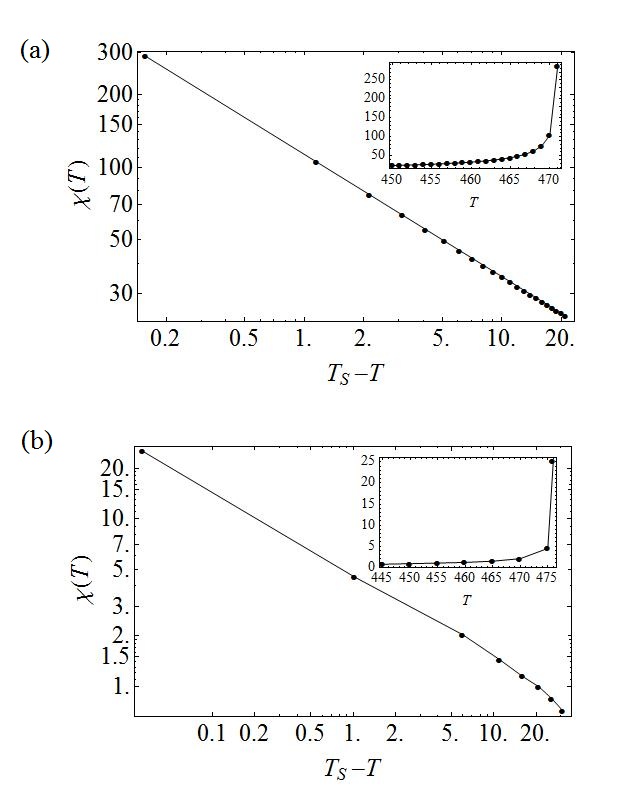}
\caption{\label{fig:gamma} Log-log plots of $\chi(T)$ vs.~$T_S-T$ for the PS model obtained from (a) the $\phi^4$ field theory and (b) simulations.  A linear plot of $\chi$ vs. $T$ is shown in the inset of both figures.}
\end{center}
\end{figure}

Since the spinodal is a critical-like point, the susceptibility $\chi=\frac{d\phi_{MS}}{dh}$ is expected to diverge as the spinodal temperature $T_s$ is approached.
We calculated  $\chi\approx \frac{\phi(h+\Delta h)-\phi(h)}{\Delta h}$ at various temperatures within the metastable region using $\Delta h=10^{-5}$.  
As the $T\rightarrow T_s$, a power law divergence of the form $\chi\sim|T_S-T|^{-\gamma}$ is observed as shown in Fig.~\ref{fig:gamma}a.  
Fitting this plot to a power law, we obtain a $T_S\approx471$~K and $\gamma\approx0.5$.  

The value of $\gamma$ can be explained if one compares it to the analogous divergence $\chi=|h_S-h|^{-1/2}$ in the Ising model.  
The coefficient of the linear term in a $\phi^4$ mapping of the Ising free energy is the field $h$, while the linear coefficient of the PS free energy is linearly dependent on temperature.  
As such, the temperature divergence in the PS model should be the same as the field divergence in the Ising model, leading to an exponent $\gamma=0.5$.  
Such must be the case as the $T$ divergence in the Ising model characterized by $\gamma=1$ only occurs by lowering $T$, whereas the spinodal in the PS model is approached by raising $T$.

\section{Mean Field Theory in the PDB Model}

While constructing a $\phi^4$-field theory would be appreciably more difficult in the PDB case, we can still obtain reasonable results for the exponent $\gamma$ with a simpler mean field theory.  
We start from the long-range Peyrard-Dauxois-Bishop model of DNA.  
Since all base pairs in the mean field model are equivalent, we first write the Hamiltonian of a single base pair,
\begin{equation}
H_{Single}\left(y,\{y_j\}\right)= V_M(y)+V_I\left(y,\{y_j\}\right).
\end{equation}
Here, 
\begin{equation}
V_M(y)=D\left(e^{-ay}-1\right)^2
\end{equation}
is the on-site Morse potential and
\begin{eqnarray*}
V_I\left(y,\{y_j\}\right)&=&\frac{K}{2R} \sum_{j}\left\{y^2-2yy_j+y_j^2\right.\\
& &\left.+ \rho e^{-\alpha y}e^{-\alpha y_j}(y^2-2yy_j+y_j^2) \right\}
\end{eqnarray*}
is the interaction potential.  
The sum is over all base pairs $y_j$ within range $R$.
The kinetic energy term has been dropped for simplicity.

In mean field, we take the limits $N\rightarrow\infty$ and $R\rightarrow N/2$.
In the Hamiltonian above, there are five types of interaction terms that depend on the separation of neighboring base pairs. 
Within the sum, these terms are proportional to $y_j$, $y_j^2$, $e^{-\alpha y_j}$, $y_j e^{-\alpha y_j}$, and $y_j^2e^{-\alpha y_j}$.
In order to have a self-consistent mean field theory, it is necessary not only that
\begin{equation}
\left<y\right>=\frac{1}{N}\sum_j y_j, \nonumber
\end{equation}
 but also
\begin{equation}
\left<y^2\right>=\frac{1}{N}\sum_j y_j^2, \nonumber
\end{equation}
\begin{equation}
\left<e^{-\alpha y}\right>=\frac{1}{N}\sum_j e^{-\alpha y_j}, \nonumber
\end{equation}
\begin{equation}
\left<ye^{-\alpha y}\right>=\frac{1}{N}\sum_j y_je^{-\alpha y_j}, \nonumber
\end{equation}
and
\begin{equation}
\left<y^2e^{-\alpha y}\right>=\frac{1}{N}\sum_jy_j^2e^{-\alpha y_j}.   \nonumber
\end{equation}
All five of these relations must hold for self-consistency.

To construct a self-consistent mean field theory, we first rewrite the mean field Hamiltonian as
\begin{eqnarray*}
H_{MF}(y)&=& D\left(e^{-ay}-1\right)^2\\
&+& K\left\{y^2-2yc_1+c_2 
+ \rho e^{-\alpha y}(y^2c_3-2yc_4+c_5) \right\}.
\end{eqnarray*}
At present, $c_1$, $c_2$, $c_3$, $c_4$, and $c_5$ will be treated as parameters that can take arbitrary values.  
In order for the theory to be self-consistent, these parameters must be chosen such that
\begin{eqnarray*}
c_1&=&\left<y\right>, \\
c_2&=&\left<y^2\right>, \\
c_3&=&\left<e^{-\alpha y}\right>,  \\
c_4&=&\left<ye^{-\alpha y}\right>, 
\end{eqnarray*}
and
\begin{equation}
c_5=\left<y^2e^{-\alpha y}\right>. 
\end{equation}

In order to compute the self-consistent values for the above parameters, we use the following procedure. 
We begin by assigning the values $c_1=0$, $c_2=0$, $c_3=1$, $c_4=0$, and $c_5=0$.  
Using these starting values, we determine a new set of parameters from the Maxwell Boltzmann probability distribution  $P(y)\propto\exp\{-\beta H_{MF}(y)\}$.  
This distribution can be normalized by dividing by the partition function
\begin{equation}
Z=\int_{-\infty}^\infty dy  \exp\{-\beta H_{MF}(y)\}.
\end{equation}
Once normalized, this distribution can be used to calculate new values for the parameters from the relation,
\begin{equation}
\left<O\right>=\int dy O P(y).\label{MBDist}
\end{equation}
Using equation (\ref{MBDist}), new values for each of the five paramerters were computed  numerically.  
This procedure was repeated until all five parameters had converged on some final value.


\begin{figure}[t]
\begin{center}
\includegraphics[width=9.cm,height=6. cm]{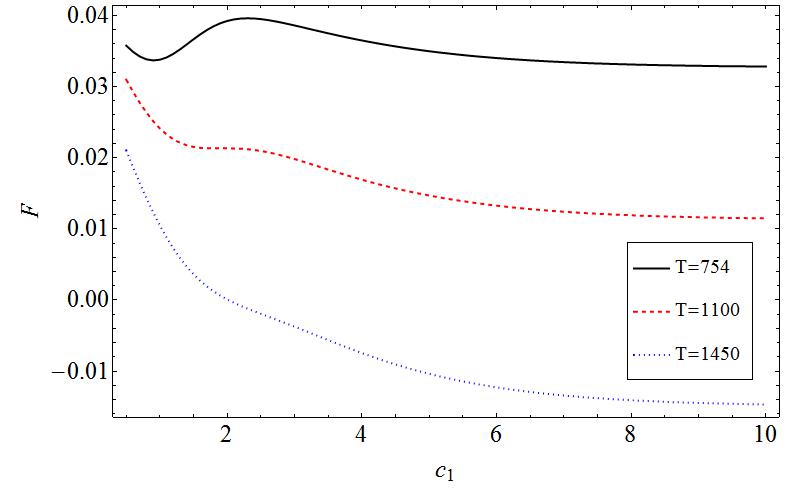}
\caption{\label{fig:fps}   (color online).  Plots of the free energy $F$ vs. $c_1$ for $T=754 K$ (black, solid), 1100 K (red, dashed), and 1450 K (blue, dotted).  For $T<1100$ K there is a well near $c_1=1$.  At $T\approx1100$ K, the well disappears.  While the presence of a well suggests metastability, this well only appears because we have restricted the system to change only along the $c_1$ axis.}
\end{center}
\end{figure}

We are particularly interested in finding the spinodal temperature.  At this temperature the metastable well disappears.
To find this temperature, we use a similar procedure to the one described above.  
This time, we solve self-consistently for the parameters $c_2$, $c_3$, $c_4$, and $c_5$ but choose the value of $c_1$.
Physically, this corresponds to allowing all the base pairs to reach equilibirum with the constraint that they must maintain a certain mean separation.
This will only be self-consistent for certain values of $c_1$.
In  Fig.~\ref{fig:fps},  we plot the free energy $F(c_1)=-k_B T \ln Z(c_1)$ as a function of $c_1$ for several temperatures.  
Here, $Z(c_1)$ is a restricted partition function, which is obtained by integrating over states with a  particular (not necessarily self-consistent) value of $c_1$,
\begin{equation}
Z(c_1)=\int_{-\infty}^\infty dy P(y).
\end{equation}
For $T<1100$ K one can see what appears to be a metastable well.
While this figure is useful for illustration purposes, it gives the false impression that the spinodal occurs at $T\approx1100$ K.
This is incorrect.  
While there is clearly a well when the free energy is projected on the $c_1$ axis, we have not considered what happens to the free energy 
when we change any of the other four parameters.  
Unless the well is a local minimum in the space of all five parameters, then the system will not be trapped at this value of $\left<y\right>$.
As it turns out, this projection of the free energy overestimates the actual spinodal temperature.

\begin{figure}[t]
\begin{center}
\includegraphics[width=9.cm,height=6. cm]{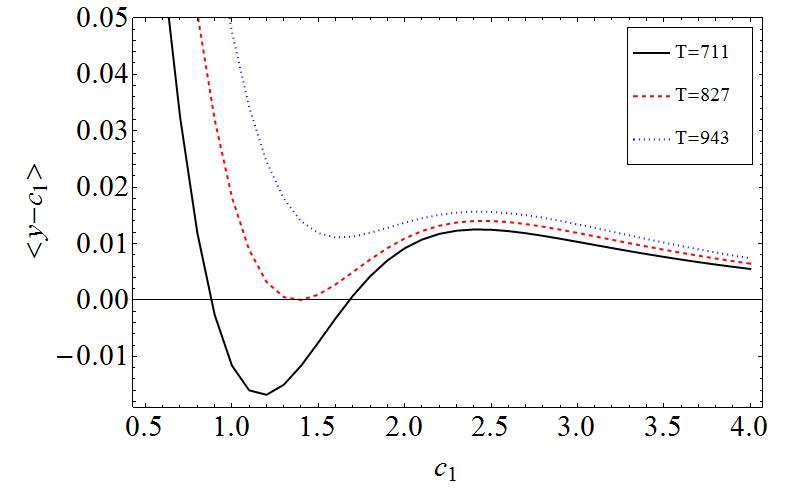}
\caption{\label{fig:scmf} (color online). Plots of $\left<y-c_1\right>$ vs. $c_1$ for $T=711$ K (black, solid), 827 K (red, dashed), and 943 K (blue, dotted).  For $T<827$ K there are two zeros that correspond to metastable and unstable fixed points. At $T=827$ K the two fixed points collide and give a spinodal.}
\end{center}
\end{figure}

To find the correct spinodal temperature, we again use a similar procedure to the one described above.  
As before, we solve self-consistently for the parameters $c_2$, $c_3$, $c_4$, and $c_5$ but leave $c_1$ as a free parameter.
We then calculate the mean separation $\left<y\right>$ as a function of $c_1$.
In  Fig.~\ref{fig:scmf},  we plot $\left<y-c_1\right>$ vs. $c_1$ for several temperatures.  
This plot was obtained by numerically integrating the Maxwell Boltzmann distribution,
\begin{equation}
\left<y-c_1\right>=\int (y-c_1)e^{-\beta H_{MF}(y)}dy.
\end{equation}
Self-consistency requires that $\left<y-c_1\right>=0$.  
Positive values of $\left<y-c_1\right>$ indicate that the mean base pair separation would be larger than $c_1$ in equilibrium as individual base pairs would drift toward larger separations. 
Negative values of $\left<y-c_1\right>$ indicate that the mean separation should be smaller than $c_1$ because base pairs would drift toward smaller values.  
The first zero in the plot is then a stable fixed point corresponding to a stable or metastable minimum.  
When a second zero appears, it represents an unstable fixed point.
At $T\approx827$ K the two zeros fuse.
Above this temperature, there are no zeros, so the system is unstable.
As such, this temperature corresponds to the limit of metastability and represents the spinodal. 
If one attempts to find $\left<y\right>$ for $T>827$ using the full self-consistent, s/he finds that the parameters will not converge and the mean separation will grow without bound.


\begin{figure}[t]
\begin{center}
\includegraphics[width=9.cm,height=10.5cm]{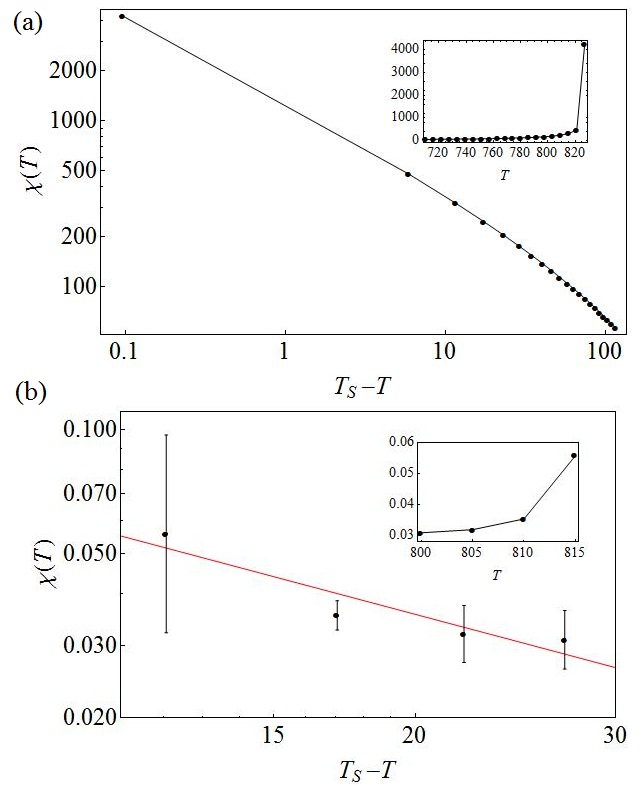}
\caption{\label{fig:pbgamma} Log-log plots of $\chi(T)$ vs.~$T_S-T$ for the PDB model obtained from (a) the mean field theory and (b) simulations.  A linear plot of $\chi$ vs. $T$ is shown in the inset of both figures.}
\end{center}
\end{figure}

Using the position of the fixed point and the definition $\chi(T)=\frac{d\left<y\right>}{dh}$, we calculated $\chi(T)$ and the exponent $\gamma$ associated with susceptibility's divergence.   
We approximate the derivative as
\begin{equation}
\left.\frac{d\left<y\right>}{dh}\right|_T\approx\frac{\left<y\right>_{h=dh}-\left<y\right>_{h=0}}{dh}
\end{equation}
with $dh= 1\times10^{-12}$.  
From the $\chi(T)$ vs.~$T$ in the inset of Fig.~\ref{fig:gamma}a, we see a sharp divergence as we raise the temperature.  
Using a two parameter fit, we observe $\gamma\rightarrow 0.5$ as $T\rightarrow T_s$, consistent with the results for the $\phi^4$-field theory of the PS model.

\section{Simulations Results}

We simulate the PDB and PS models using Brownian dynamics (BD) and the Metropolis Monte Carlo (MC) algorithm, respectively.  In BD, the system evolves via a Langevin equation 
\begin{equation}
m\ddot{y_i}=-\nabla V(y_i)-\gamma\dot{y}+\eta(t)
\end{equation}
where the noise $\eta(t)$ is random Gaussian with $\left<\eta(t)\right>=0$ and $\left<\eta(t)\eta(t^\prime)\right>=2\gamma k_B T \delta(t-t^\prime)$.  
In time units of $\tau=1.018\times10^{-14}$\,s, we use a time step of $0.25\,\tau$ and a damping constant $\gamma=10^{-4}~\tau^{-1}$.
In the MC simulations, random spins are flipped and the change in energy between the original and final states is calculated.  
Negative changes in energy are always accepted, while positive changes are accepted with probability $\exp(-\Delta E/k_B T)$, where $\Delta E$ is the change in energy.  
Both simulations use periodic boundary conditions for simplicity.
In the next two subsections, we describe nucleation in homogeneous systems while in the third subsection we consider nucleation in heterogeneous systems.

\subsection{Evidence for Metastability}

\begin{figure}[t]
\begin{center}
\includegraphics[width=9.cm,height=7. cm]{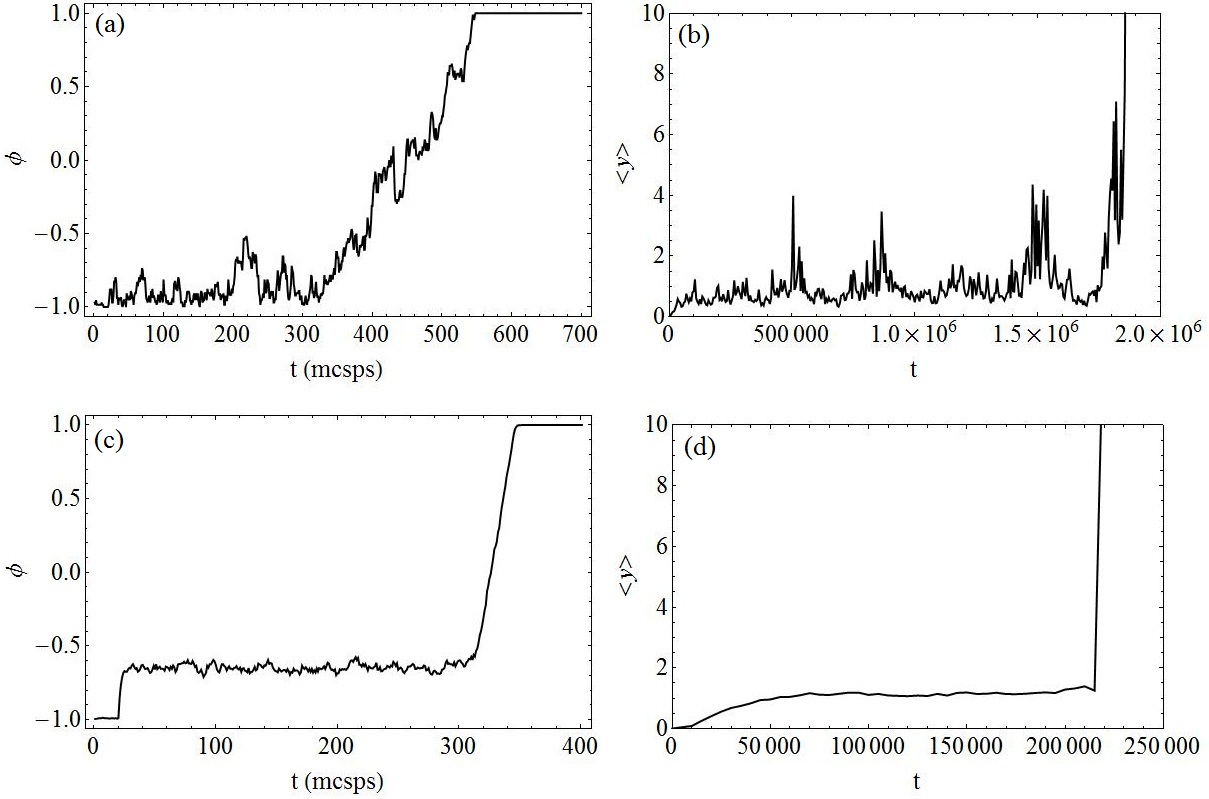}
\caption{\label{fig:MS}   Metastability and Nucleation.  Plots are shown for the time evolution  of (a) the $N=128$ PS model with $R=1$ quenched to $T=365$ K, (b) the $N=128$ PDB model with $R=1$ quenched 
to $T=380$ K, (c) the $N=4096$ PS model with $R=205$ quenched to $T=470$ K, and  (d) the $N=4096$ PDB model with $R=205$ quenched to $T=800$ K.  }
\end{center}
\end{figure}

We plot the time evolution of both models for ranges $R=1$ and $R=205$ in Fig.~\ref{fig:MS}.  
For the PDB model, we monitor the growth of the mean separation of the strands $\left<y\right>=\sum y_i/N$, 
while for the PS model we plot the ``magnetization" $\left<\sigma\right>=\sum \sigma_i/N$.  
Each run is quenched instantaneously from $T=300$~K to a higher temperature listed in the caption.  
Each temperature is chosen to give a metastable lifetime that is long enough to clearly demonstrate metastability but short enough to have a reasonably fast run time.

For both ranges in the PS model, we observe a period where the magnetization $M$ stabilizes before growing sharply.
Growth is visibly sharper for systems with long-range interactions. 
Within the time allotted for the runs, the system never returned to the bonded state after complete separation.
Stability prior to spontaneous growth is the hallmark of nucleation out of a metastable state.
In addition to these signatures, we observe the formation of a droplet (see below) that grows into the stable phase much like those observed in real DNA.

In the PDB model,  we observe similar stability and growth of the mean bp separation $\left<y\right>$ for both ranges, again with sharper growth for the long-range system.
Unlike the PS model, the NN PDB model will frequently rebind even after the strands have completely separated.
Bases in real DNA bubbles sample a large three-dimensional phase space, while bases in PDB bubbles can only move in one dimension.
For this reason, bases in PDB bubbles are more likely to find and rebind with their complementary pair than bases real DNA bubbles which must search a larger space.
The result is analogous to a one-dimensional random walker which will inevitably return to its starting position after some finite time.
Recombination of the strands was not observed in runs of the LR PDB model.  
While it is possible that one might observe recombination of strands in a LR system given enough time, a more likely explanation is that the LR interactions create a greater entropic barrier that suppresses the likelihood of reforming the double strand.

It should be noted that both the PS and PDB models exhibit noticibily smaller fluctuations when the interactions are long-range.  
This is expected since increasing the interaction range takes the system closer to its mean field approximation in which there are no fluctuations.
Practically speaking, reducing the size of fluctuations is beneficial since one need not wait very long to observe that the system has reached metastable equilibrium.
This is particularly useful in long-range systems, where the simulation speed is inherently slower.
In contrast, the NN PDB model exhibits fluctutations large enough that one must observe long runs in order to clearly see that the system has reached metastable equilibrium.

\begin{figure}[t]
\begin{center}
\includegraphics[width=9.cm,height=9. cm]{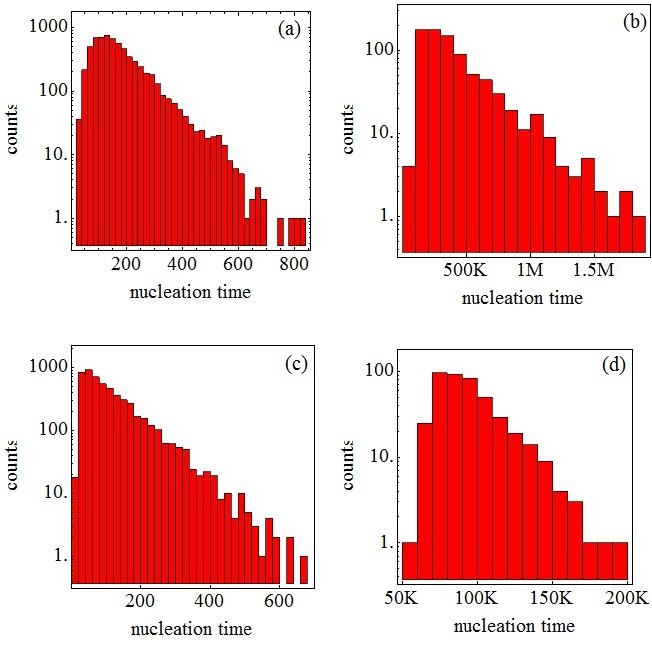}
\caption{\label{fig:NTH}   Histogram of nucleation times for (a) the PS model with $N=128$ and $R=1$ quenched to $T=370$ K, (b) the PDB model with $N=128$ and $R=1$ quenched to $T=400$ K, 
(c) the PS model with $N=4096$ and $R=205$ quenched to $T=470$ K, and (d) the PDB model with $N=1024$ and $R=50$ quenched to $T=780$ K.  
}
\end{center}
\end{figure}

Since nucleation is an activated process governed by a constant rate, it should follow Poisson statisics.  
We determine the nucleation rate for both models and ranges by measuring the nucleation time of an ensemble of systems run with a different random noise.  
We define the nucleation time  as the time at which either the mean magnetization in the PS model or the mean separation in the PDB model is greater than some threshold.
The threshold values were chosen to be sufficiently larger than metastable fluctuations and are listed in the caption of Figs.~\ref{fig:NTH}, in which we plot on a log-scale histograms of the nucleation times.  
In both long- and short-range models, we observe exponential decay for large times, indicating that systems reach the stable state at a constant rate.
Strand separation at a constant rate is consistent with the notion that this DNA melting occurs via nucleation.

\subsection{Divergence of the Susceptibility}

We calculated the susceptibility at various temperatures by measuring the fluctuations in the PS model,
\begin{equation}
\chi_T=\frac{\left<\phi^2\right>_t-\left<\phi\right>^2_t}{T},
\end{equation}
and the PDB model,
\begin{equation}
\chi_T=\frac{\left<\left<y\right>_x^2\right>_t-\left<\left<y\right>_x\right>^2_t}{T}.
\end{equation}
Here, $\left<...\right>_t$ and $\left<...\right>_x$  signify time and spatial averages, respectively.  

As one approaches the spinodal, the system nucleates quickly.
To suppress nucleation, we used very large interaction ranges.  
We chose $N=8193$ and $R=4096$ for the PDB model and $N=262144$ and $R=13107$ for the PS model.
These larger system sizes necessarily take a long time to run.
For the PDB model, which even for smaller rangges has fairly long run times, this makes obtaining accurate statistics difficult.
This difficulty is compounded by the fact that the spinodal is a critical-like point at which the correlation time blows up.
This critical slowing down results in fairly large error ranges for $\chi$.

We plotted $\chi$ vs $T$ simulation results in the insets of Fig.~\ref{fig:gamma}b and  Fig.~\ref{fig:pbgamma}b. 
A divergeance appears as $T$ approaches the spinodal temperature $T_s$.  
This divergence appears power law when plotted on a log-log scale vs.~$Ts-T$ (Fig.~\ref{fig:gamma}b and Fig.~\ref{fig:pbgamma}b).  
Using a power-law fit, we calculated the susceptiblity exponents 
$\gamma\approx0.7\pm0.2$ and $\gamma\approx0.50\pm0.01$
in the PDB and PS models, respectively.  
These exponents are consistent with our earlier theoretical results.

\subsection{Droplets and Growth Modes}
 
\begin{figure}[t]
\begin{center}
\includegraphics[width=8.5cm,height=9. cm]{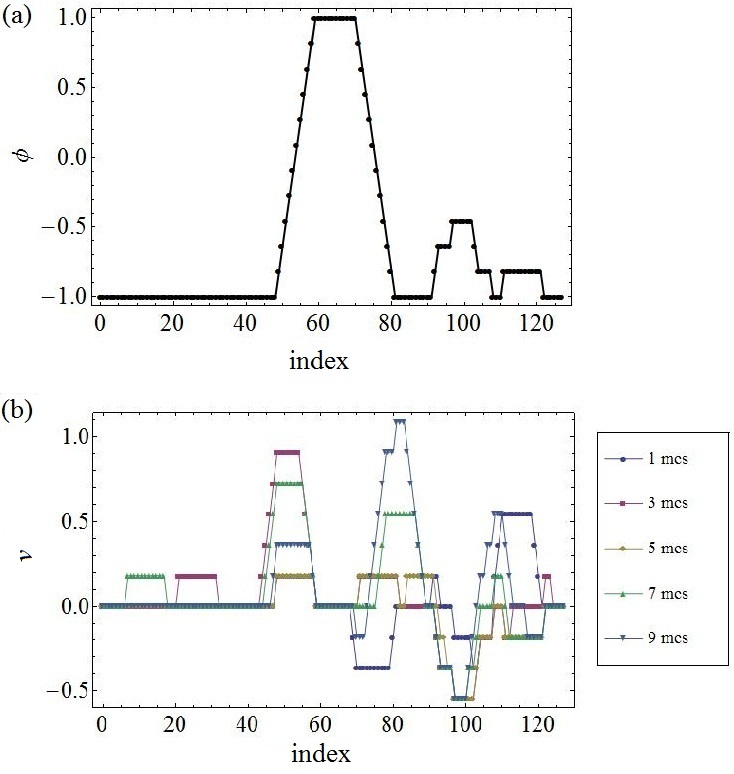}
\caption{\label{fig:PSNN}  (a)  Droplet profile and (b) initial growth in the PS model for $N=128$, $R=1$ and $T=365$ K.  The system uses a coarse-graining range $L_{CG}=5$.}
\end{center}
\end{figure} 
\begin{figure}[t]
\begin{center}
\includegraphics[width=8.5cm,height=9. cm]{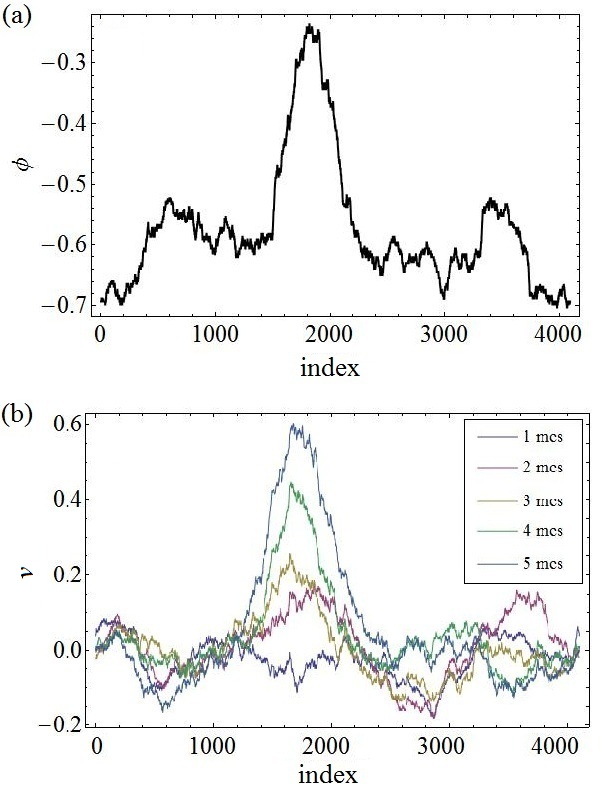}
\caption{\label{fig:PSLR}   (a) Droplet profile and (b) initial growth in the PS model for $N=4096$, $R=205$ and $T=470$ K.  The system uses a coarse-graining range $L_{CG}=205$.}
\end{center}
\end{figure}
\begin{figure}[t]
\begin{center}
\includegraphics[width=8.5cm,height=9. cm]{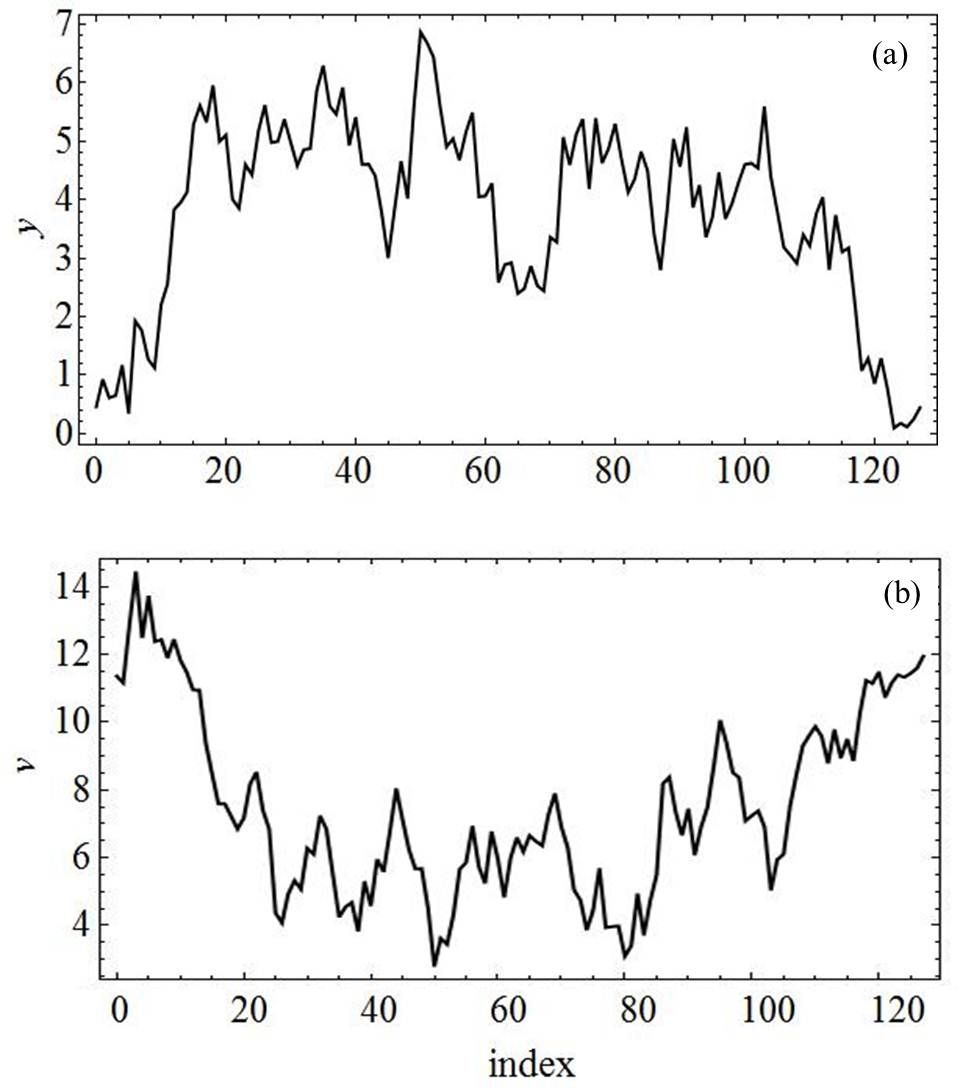}
\caption{\label{fig:PBNN}  (a)  Droplet profile and (b) initial growth in the PDB model for $N=128$, $R=1$ and $T=380$ K.  
}
\end{center}
\end{figure}
\begin{figure}[t]
\begin{center}
\includegraphics[width=9.cm,height=8. cm]{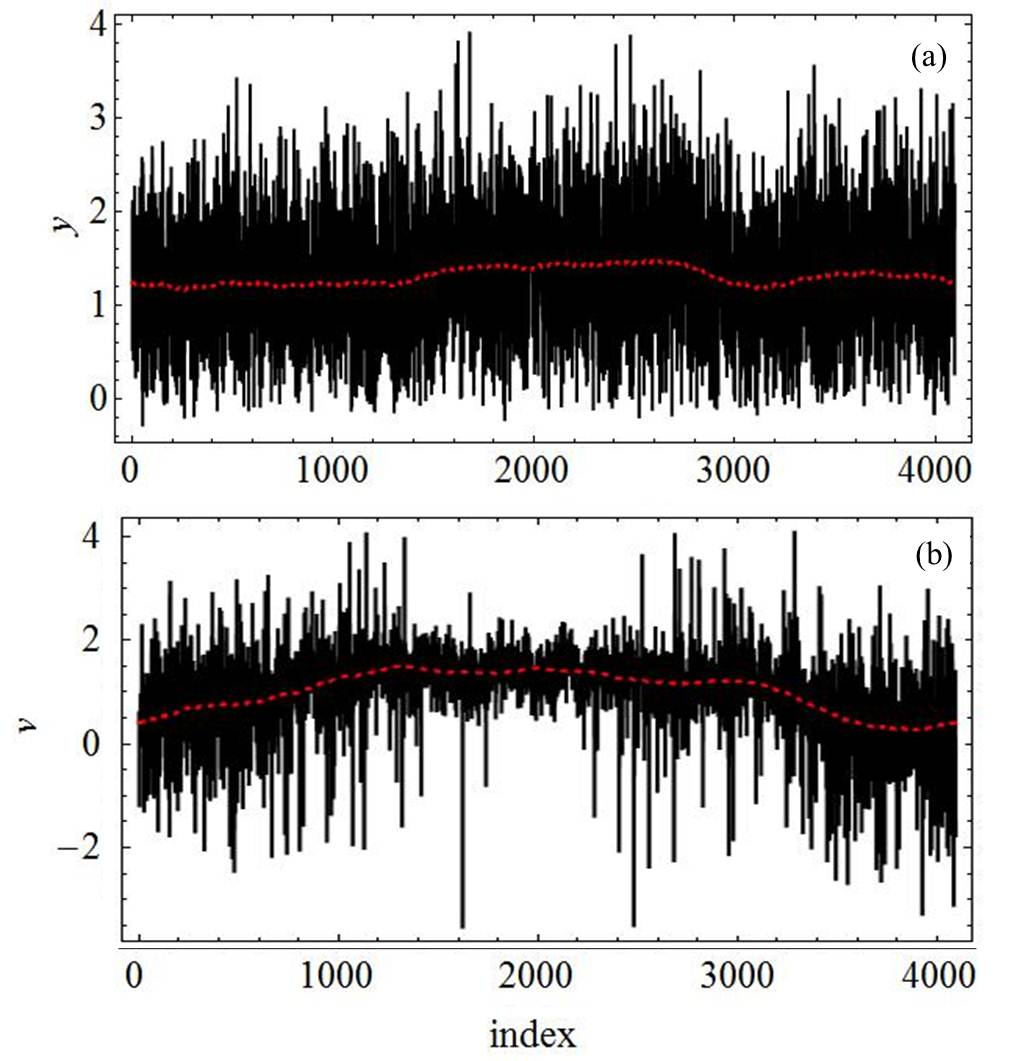}
\caption{\label{fig:PBLR} (a)  Droplet profile and (b)  initial growth in the PDB model for $N=4096$, $R=205$ and $T=800$ K.  The droplet is shown in black.  The red dashed curves are coarse-grained plots of the droplet and growth mode where the base pair separation has been averages over a coarse-graining length $L_{CG}=205$.}
\end{center}

\end{figure}
The critical droplet sits atop a saddle point hill in a free energy landscape.
At this saddle point, the system has an equal probability of nucleating or returning to the metastable well.  
To determine the shape of the critical configuration, we first run our system until the strands have clearly separated. 
We then make 20 copies of the system and rerun them using the same initial conditions and random number sequence until some intervention time $t_I$.  
At this time, each copy of the system is given a new sequence of random numbers that is different from each of the others.  
We then run the copies with the new random number sequences and calculate the percentage that still separate.  
The critical droplet is the configuration that has approximately equal likelihood of growing into the stable phase or returning to the metastable well.

In Figs. \ref{fig:PSNN}a and  \ref{fig:PSLR}a, we plot the critical droplet profiles for the PS model with ranges $R=1$ and $R=205$, respectively.
These profiles were taken at nucleation times obtained using the intervention procedure described previously.  
For easier viewing, each profile has been coarse-grained by averaging over all bps within a length $L_{CG}$,
\begin{equation}
\phi_{CG,i}= \sum_{i-L_{CG}}^{i+L_{CG}}\phi_i
\label{cg}
\end{equation}
where $\phi$ is replaced by $\sigma$.
The coarse-graining length $L_{CG}$ is listed in each figure.
Consistent with both our theoretical results and results of previous studies \cite{Langer,Unger,Unger2, Gould}, we obtain compact, large amplitude  droplets for $R=1$ and temperatures near the melting temperature, and diffuse, small amplitude droplets for $R=205$ and temperatures near our theoretical prediction for the spinodal (note the change in scale).  
In addition to the droplet profiles, the growth modes were obtained by subtracting the critical profile from profiles at later times.  
The NN model shows an initial growth that is peaked at the surface of the droplet while the LR model give rise to growth modes that are peaked at the center of the droplet.  
These results are consistent with the notion that nucleation in the lower temperature NN system occurs near a coexistence curve, while the LR system at a higher temperature nucleates close to a spinodal~\cite{Langer,Unger,Unger2, Gould}.

In Figs. \ref{fig:PBNN}a and  \ref{fig:PBLR}a, we plot the critical droplet profiles for the PDB model with $R=1$ and $R=205$, respectively. 
As before, critical profiles were taken at nucleation times determined by intervention.  
The Morse well becomes flat in the range $1\leq y\leq 2$, so bps with separations greater than 2 can be considered open.
The NN critical profile exhibits a compact open region with very large bp separations, i.e. $y_i>10$. 
Unlike the PS model, the region spans almost the entire length of the droplet (see discussion below).
In contrast to the NN system, the LR PDB model produces diffuse critical droplets that differ little from the metastable background.
These droplets are similar to those found using the LR PS model.
Using Eq. \ref{cg} with $\phi$ replaced by $y$ and $L_{CG}=205$, we coarse-grained the LR droplet profile to obtain a sharper image of its shape (red curve in Fig.  \ref{fig:PBLR}).  
As with the PS model, we found growth modes for the PDB model by subtracting the critical profile from profiles at later times.  
The results are displayed in Figs. \ref{fig:PBNN} and  \ref{fig:PBLR}b for various times after the critical droplet.
Though noisy, one can easily see that growth occurs at the droplet's edge in the NN system and at the droplet center in the LR system. 
These growth modes are consistent with those found in the PS models and those predicted by nucleation theory~\cite{Langer,Unger,Unger2, Gould}..

Classical nucleation theory predicts compact droplets greater than some critical size will grow from their edges to bring the system into the stable phase.  
This does not appear to be the case for NN PDB droplets.  
As noted earlier, the NN PDB droplet shown in Fig. \ref{fig:PBNN} has the vast majority of 128 bps already open, with only about 20 bps still bonded.  
If the simulation is run using the same parameters but with different random noise, similar results are obtained, again with roughly only 20 bps remaining bonded in the critical configuration.
Moreover, if we increase the system size to $N=256$ but keep the temperature, range, and other parameters fixed, we find that the system again transitions to the unbound state with a critical configuration that has the vast majority of its bps unbound and roughly 20 still in the bonded state. 
Since the critical droplet grows with the size of the system, the NN model does not appear to be undergoing classical nucleation.
It is possible this is due to a finite size effect, but it is difficult to ascertain whether or not this is the case.
In principle, we should be able to increase the system size until finite size effects go away, but in practice this is quite difficult.
As mentioned earlier, the large flutuations that arise in the NN model make it difficult to see when metastable equilibrium has been reached, so one must lower the temperature to produce longer metastable lifetimes.
Increasing the system size makes these already very long runs even longer.
Furthermore, the range of interaction is much smaller than the system size, so it seems unlikely that this discrepency can be explained purely by finite size effects. 

If finite size effects do not cause the discrepency between NN PDB droplets and those predicted by classical nucleation theory, then melting in the NN PDB model likely occurs via some mechanism other than nucleation. 
There is evidence to support this hypothesis.
First, while the transition is sharp with a peaked in the heat capacity near the melting temperature, it has not yet been shown to be a true phase transition~\cite{Dauxois,Dauxois2}.
This is in contrast to the NN PS model, which includes the entropy of three-dimensional unbound loops to obtain a true phase transition when $c>2$~\cite{Kafri}.  
Second, our simulations of the NN PDB model show recombination of the strands even after complete separation, suggesting again that melting is more likely a fluctuation than a first-order phase transition.
Furthermore, Van Hove argued that phase transitions do not occur in one-dimension for systems with finite-range interactions~\cite{VanHove}.
Strictly speaking, his argument does not apply here since it assumes the absence of external potentials.  
Indeed, one can easily see that one-dimensional phase transitions do occur by considering a one-dimensional Ginzburg-Landau model with the $\phi^4$ polynomial expansion terms treated as an external potential.
Unlike the Ginzburg-Landau model, the NN PDB does not have a second energetic well to represent the bound state.
As shown in Section IV, the mean field PDB model does exhibit an entropic well caused by the nonlinear term in the coupling interaction, but the mean field model clearly contains LR interactions. 
It seems likely that the nonlinear coupling term is insufficient to induce a phase transition via nucleation when the interaction range is NN.
In short, while the melting transition in the NN PDB model seems to be an activated process induced by a fluctuation, it does not appear to transition via true nucleation.
As one increases the range of interaction, the transition is sharper and can be more accurately described by nucleation.


\subsection{Heterogeneous Nucleation}
\begin{figure}[t]
\begin{center}
\includegraphics[width=8.8cm,height=6.8 cm]{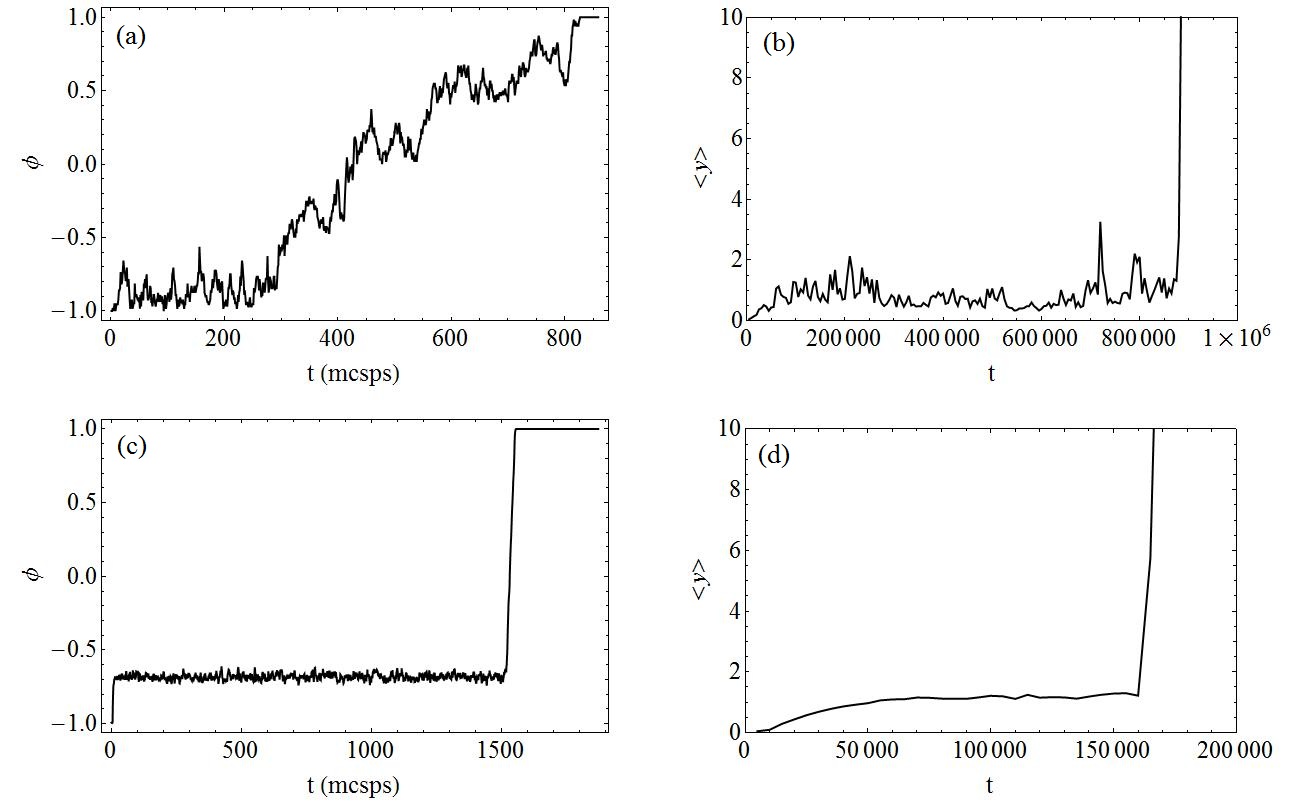}
\caption{\label{fig:heteroms}   Metastability for Heterogeneous Sequences.  Plots are shown for the time evolution  of (a) the $N=128$ PS model with $R=1$ quenched to $T=365$ K, (b) the $N=128$ PDB model with $R=1$ quenched to $T=380$ K, (c) the $N=4096$ PS model with $R=205$ quenched to $T=465$~K, and  (d) the $N=4096$ PDB model with $R=205$ quenched to $T=805$ K.  }
\end{center}
\end{figure}

\begin{figure}[t]
\begin{center}
\includegraphics[width=8.5cm,height=13. cm]{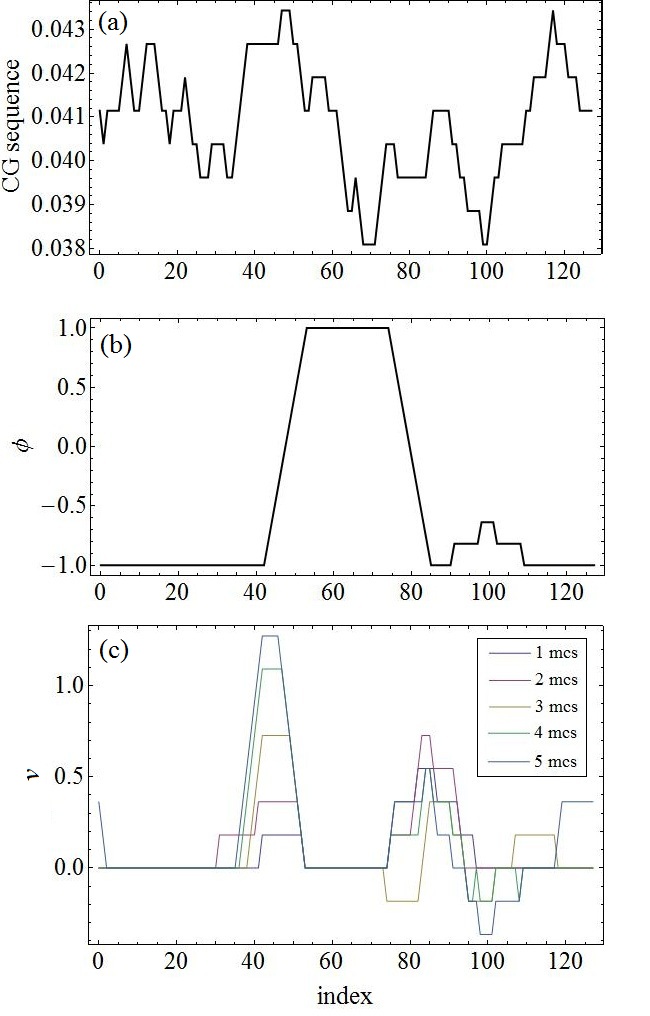}
\caption{\label{fig:heterpsnn}  Coarse-grained dissociation energy, droplet profile, and initial growth in the PS model for $N=128$, $R=1$ and $T=365$ K.  The system uses a coarse-graining range $L_{CG}=5$.}
\end{center}
\end{figure}

\begin{figure}[t]
\begin{center}
\includegraphics[width=8.5cm,height=13. cm]{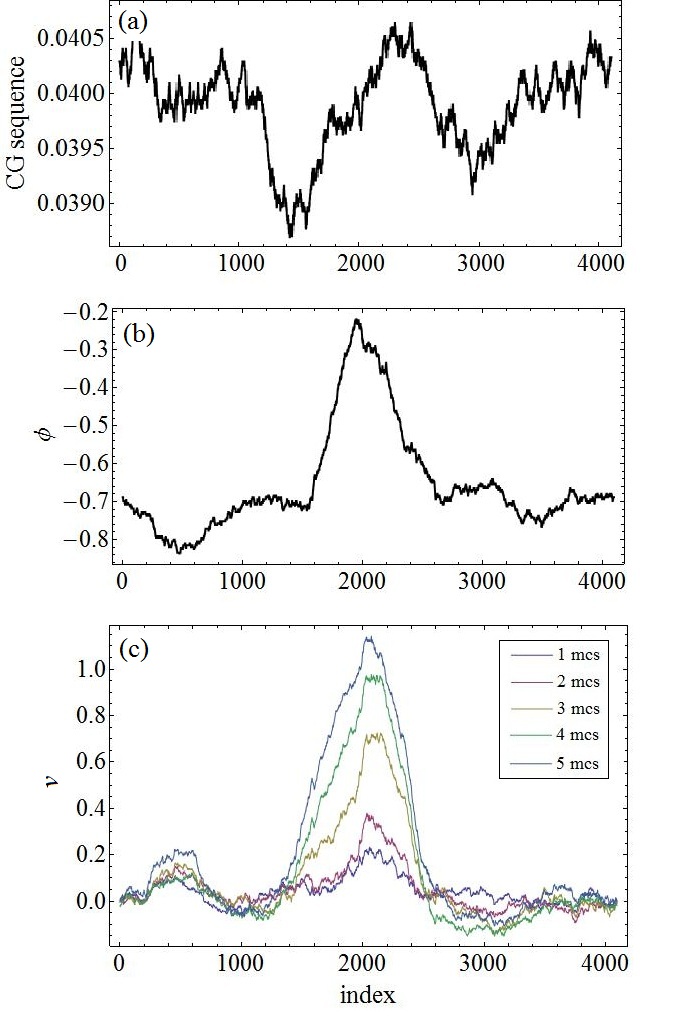}
\caption{\label{fig:heteropslr} Coarse-grained dissociation energy, droplet profile, and initial growth in the PS model for $N=4096$, $R=205$ and $T=465$ K.  The system uses a coarse-graining range $L_{CG}=205$.}
\end{center}
\end{figure}

\begin{figure}[t]
\begin{center}
\includegraphics[width=8.5cm,height=13. cm]{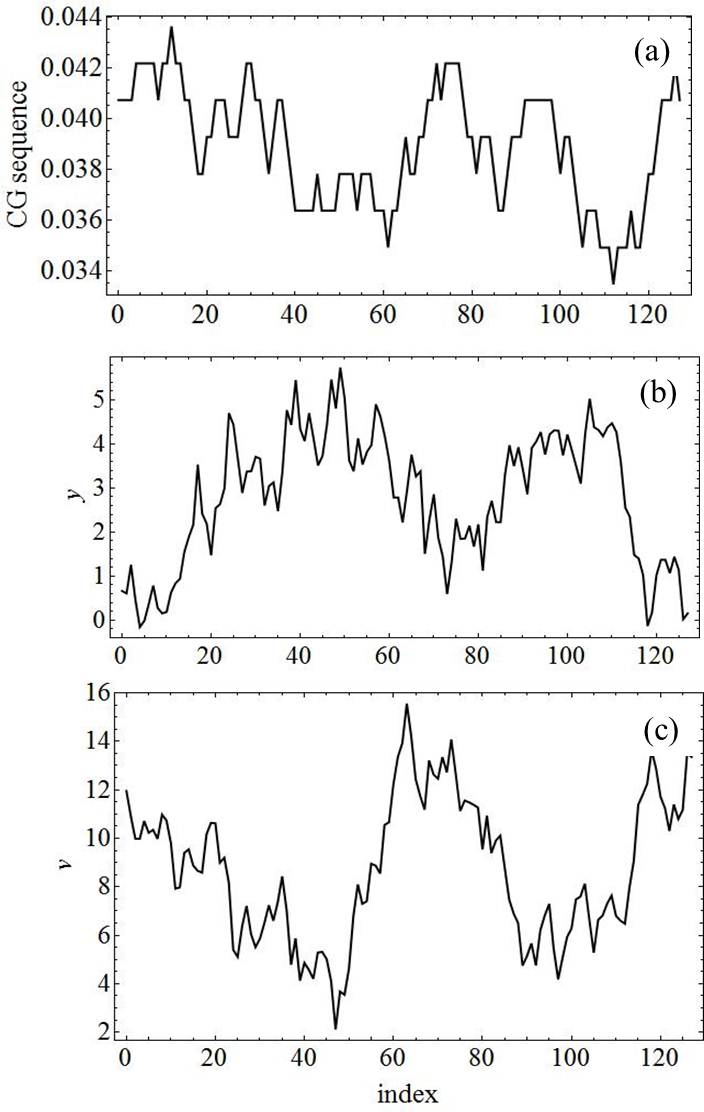}
\caption{\label{fig:heterpbnn} Coarse-grained dissociation energy, droplet profile, and initial growth in the PS model for $N=128$, $R=1$ and $T=380$ K. }
\end{center}
\end{figure}
\begin{figure}[t]
\begin{center}
\includegraphics[width=8.5cm,height=13. cm]{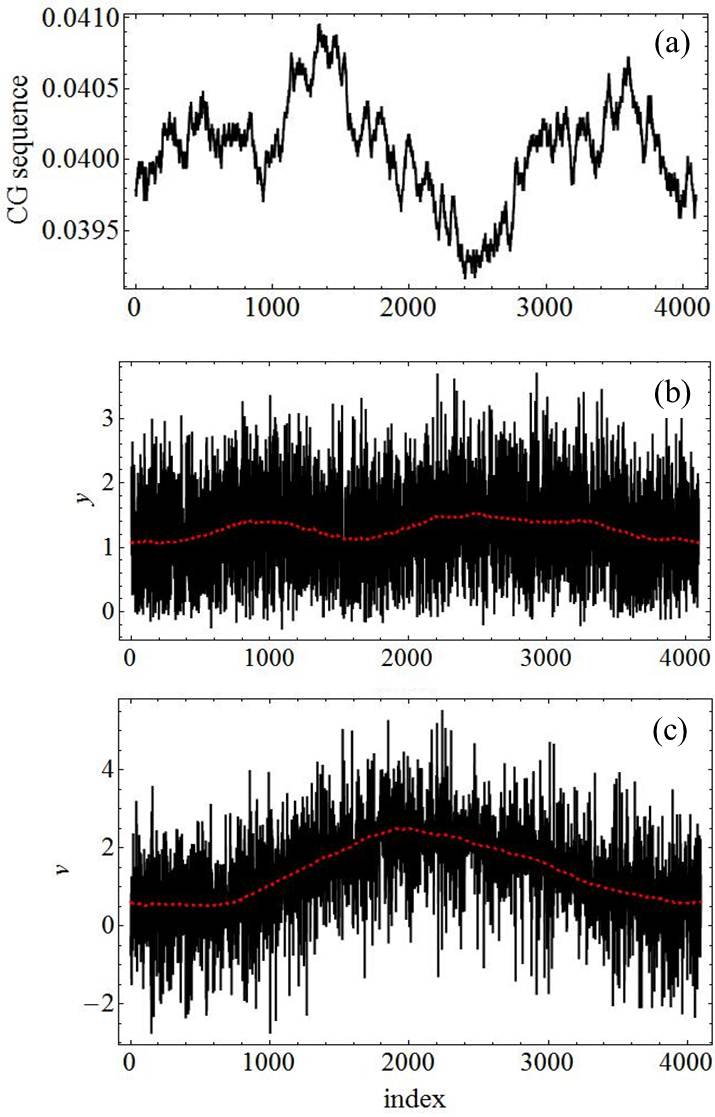}
\caption{\label{fig:heterpblr} Coarse-grained dissociation energy, droplet profile, and initial growth in the PS model for $N=4096$, $R=205$ and $T=850$ K.  The red dashed curves are coarse-grained plots of the droplet and growth mode where the base pair separation has been averages over a coarse-graining length $L_{CG}=205$..}
\end{center}
\end{figure}

Biological DNA is heterogeneous, so if our results are to have biological significance they must hold for heterogeneous systems.
To test whether or not this is the case, we simulated the PS and PDB models for randomly generated sequences.  
Each bp was assigned a dissociation energy of either $D_{AT}$ or $D_{CG}$ with equal likelihood.
As with our homogeneous results, we again monitor the time evolution of both models by plotting $\left<y\right>$ and $\left<\sigma\right>$ for $R=1$ and $R=205$ in Fig.~\ref{fig:heteroms}.
As before, each run is quenched from $T=300$ K to a higher temperature whose value is chosen to give a reasonable run time.
The value of the quenched $T$ is given in the caption of each figure.
As with homogeneous systems, we see a brief period of stability prior to spontaneous growth in both models for both ranges, indicating that the heterogeneities do not eliminate metastability.

For each of the runs in Fig.~\ref{fig:heteroms}, we determine the nucleation time using the intervention procedure described earlier.  
For each model and range, we again plot the critical droplet profile and growth modes (Figs. \ref{fig:heterpsnn}-\ref{fig:heterpblr}).
In the top of each figure, we plot the coarse-grained dissociation energy so that one can easily visualize how binding energy changes across the sequence.
For each site $i$, the coarse-grained dissociation energy $D_{CG,i}$ is given by 
\begin{equation}
D_{CG,i}= \sum_{i-L_{CG}}^{i+L_{CG}}D_i.
\end{equation}
Though the results are nosier than the homogeneous case, we still observe that for both the PS and PDB NN models where quenches are restricted to be near the coexistence curve, melting is initiated by compact large-amplitude droplets.
As before, the critical drolet of the NN PDB model spans most of the system.
The growth modes of both NN models feature peaks at the surface of the droplet.
In both models, LR interactions allow the system to reach metastability at higher temperatures.
This brings the system closer to an apparent pseudospinodal, which gives rise to diffuse critical droplets whose amplitude is close to the metastable background.
Both LR systems grow from the center of the droplet.
These results suggest that  the inclusion of random heterogeneities does not have a major effect on the qualitative features of the transition.

The formation and growth of a critical droplet will differ from seqeunce to sequence and even from run to run for a particular sequence of bps. 
Still, it is worthwhile to analyze the particular pathways that the heterogenous systems reported here have taken.
First, one will note that  droplets of both models tend to be peaked around areas where the binding strength is weak.  
For example, the NN PS model features a dip in the binding energy around bp index 70, which also corresponds to the peak of the droplet.
Similarly, the critical droplet in NN PS model features two peaks around bp indexes 50 and 100, which correspond to dips in the binding energy.
Particularly noticible is the dip in the critical droplet near bp index 75 that appears between these two peaks.
This dip corresponds to a stronger than average binding strength.
A similar dip in the LR PDB droplet occurs between bp indexes 1000 and 2000, at which there is a maximum in the binding energy of the strands.
While the criticial droplet of the LR PS model shown in Fig. \ref{fig:heteropslr} does not correspond to a large dip in the binding energy, the growth mode appears to develop a shoulder on the left side of the droplet at which there is a prominent dip in the binding energy.  
These results suggest that while modest amounts of inheterogeneities do not completely distort the qualitative character of the transition, they do make melting pathways at weaker sites more probable than pathways through strongly bound bps.
This would imply that heterogeneities cause only minor changes in the droplet structure but cause appreciable changes in the nucleation rate.

\section{Summary and Discussion}

We have extended the PS and PDB models of DNA to include variable-range interactions and have observed nucleation in the both the original and long-range models.  
In the PS model, we constructed a $\phi^4$-field theory by enumerating all possible spin configurations of a coarse grained region, calculating the average entropy of the region due to both loops and 
coarsening, and assuming that this average was equivalent to the total entropic contribution from the region.
While this field theory is not expected to produce accurate results near the coexistence curve where the coarse-graining size is smaller than the largest loops, it is useful in calculating droplet profiles and growth modes at higher temperatures where the droplets become more diffuse.  
In addition, the field theory predicts a divergent susceptibility near the spinodal with a pseudocritical exponent $\gamma\approx 0.5$.  

We constructed a self-consistent mean field theory of the PDB model from which we calculated $\chi(T)$ and observed a divergence as $T\rightarrow T_s$ characterized by the spinodal exponent $\gamma\approx 0.5$.

Simulations of both models for a variety of ranges show some  signature characteristics of nucleation including metastability prior to growth and a constant transition rate. 
Critical droplets and growth modes were measured both near the coexistence curve and near the spinodal in systems with short and long range interactions, respectively.  
We find that the NN PS model produces compact large amplitude droplets similar to those of classical nucleation theory, while the NN PDB model produces droplets that span nearly the entire system size.
Both NN models appear to grow from the droplet surface.
Both the PDB and PS models exhibit diffuse small amplitude droplets when given LR interactions and quenched near the theoretically-predicted spinodal.
As expected, we find compact classical droplets growing predominantly at their surface and diffuse spinodal droplets growing from their center.  
The presence of heterogeneous sequence does  not appear to appreciably alter the above results.
By measuring the fluctuations, we calculated $\chi(T)$ and 
observed a divergence at large $T$ characterized by the pseudo critical exponent consistent with our theoretical  results.

The consistency of the critical exponents in both models suggests that they may be in the same universality class for spinodals.
By measuring the susceptibility near the melting temperature, it may be possible to experimentally determine how accurate a mean field depiction of DNA is and whether or not long-range interactions 
play a significant role in determining the qualitative character of nucleation observed.  In particular, the presence or absence of long-range interactions may determine the most effective way for 
biological enzymes to mechanically denature DNA locally, and an understanding of the qualtitative difference between the long and short range systems may be valuable to researchers experimenting on these enzymes.

\section{Acknowledgements}

We would like to thank Harvey Gould, Aaron Schweiger, and Rachele Dominguez for useful discussions and in particular Kipton Barros for his help with simulations and discussions of the field theoretic
versions of the model.  We also thank both the UNCF-Merck Graduate Dissertation Fellowship and the DOE for their financial support.


\end{document}